\documentclass[11pt,a4paper]{article}
\usepackage{amsmath}
\usepackage{amsfonts}
\usepackage{amssymb}
\usepackage{color}
\usepackage{graphicx}
\usepackage[all]{xy}
\usepackage{color}
\usepackage{authblk}
\usepackage[T1]{fontenc}
\usepackage{lmodern}
\usepackage[utf8]{inputenc}
\usepackage{graphicx}

\title{{The Kantowski-Sachs quantum model with stiff matter fluid}}

\author[1]{F. G. Alvarenga\thanks{e-mail: \texttt{flavio.alvarenga@ufes.br}}}
\author[2]{R. Fracalossi\thanks{e-mail: \texttt{rfracalossi@gmail.com}}}
\author[2]{R. C. Freitas\thanks{e-mail: \texttt{rodolfo.camargo@pq.cnpq.br}}}
\author[2]{S. V. B. Gon\c{c}alves\thanks{e-mail: \texttt{sergio.vitorino@pq.cnpq.br}}}
\affil[1]{Departamento de Ci\^encias Naturais, CEUNES, Universidade Federal do Esp\' \i rito Santo, CEP 29933-415, S\~ao Mateus, ES, Brazil.}
\affil[2]{Departamento de F\' \i sica, CCE, Universidade Federal do Esp\' \i rito Santo, CEP 29075-910, Vit\'oria, ES, Brazil.}

\date{}
\begin{document}
\maketitle

\begin{abstract}
In this paper we study the quantum cosmological Kantowski-Sachs model and solve the Wheeler-DeWitt equation in minisuperspace to obtain the wave function of the corresponding universe. The perfect fluid is described by the Schutz's canonical formalism, which allows to attribute dynamical degrees of freedom to matter. The time is introduced phenomenologically using the fluid's degrees of freedom. In particular, we adopt a stiff matter fluid. The viability of this model is analyzed and discussed. 
\end{abstract}


\section{Introduction}
At first sight it may seem meaningless to study quantum cosmological models \cite{Halliwell,Moniz,Wiltshire,Nelson} taking into account only an absolutely finite number of degrees of freedom, i.e., a minisuperspace. In general, the idea is to restrict the number of degrees of freedom of the metric to a finite number, by imposing determined symmetries. As a consequence we have that the quantum constraints become easier to handle, when written in the usual variables; since we allow fewer metric variations when getting the equations of motion from the action, we may get more classical solutions, some spurious ones being unstable; from both semiclassical and quantum point of view, the results may not be equivalent to appropriate sectors of the full theory. Despite this, the construction of a quantum theory of gravity must be related, among other things, to obtaining the solution to the Wheeler-DeWit equation in the superspace. Hence, to exclude infinite degrees of freedom does not seem to be a reasonable procedure. However, the quantization in minisuperspace is still very useful for other reasons. Minisuperspace quantum models are very simple systems, without much of symmetry of general relativity. They provide non-trivial examples where we can apply the ideas of quantum gravity and mathematical techniques. They are still relevant for the discussion of major conceptual issues of quantum gravity such as the problem of time and the invariance under time reparametrization of classical general relativity and the physical interpretation of the Hamiltonian constraint. In some cases, a time coordinate may be identified with the space volume, which is a growing function in an expanding universe. But, all these attempts have limited applications until now, and the problem of time in quantum gravity and quantum cosmology remains an unsolved puzzle \cite{Simeonelivro,Simeone2,Anderson,Isham,Magain}. Even sometimes working with only the existence of the gravitational field, quantum aspects can be consistently verified or at least induced. This is even more evident in the study of quantum properties of black holes as well as the use of the known quantization processes to the universe as a whole.

Focusing only in minisuperspace we can ask what are the degrees of freedom responsible for ensuring certain fundamental aspects of the desired description of the cosmological model. Only after verifying these aspects we can take into account the thermodynamic aspects of the problem and the presence of other fields. However, the use of metrics with additional degrees of freedom to those found in FRW metric can be deeply complex, especially from a mathematical point of view. As a result, models that represent inhomogeneous and anisotropic universes are largely unexplored in comparison to the FRW. We can cite Bianchi models, for example \cite{Ellis}. These cosmological models have a limited number of analytical solutions, and to obtain numerical solutions for the Wheeler-DeWitt equation can be a complex process.

In this work, we study the quantization of minisuperspace models within the context of the Kantowski-Sachs metric, and we describe matter as a perfect fluid with an arbitrary barotropic equation of state $p = \alpha\rho$. Therefore, we will adopt Schutz's canonical formalism \cite{Schutz2,Schutz1}, which describes a relativistic fluid interacting with the gravitational field. This description is essentially semiclassical \cite{Kie}, but it has the advantage of assigning dynamical degrees of freedom connected with the matter, which can naturally be identified with time, leading to a well-defined Hilbert space structure. In particular, the stiff matter fluid is used. The importance of the stiff matter in cosmological models is discussed in \cite{Germanomr}.

The Kantowski-Sachs space-time \cite{Kantowski-Sachs} is a cosmological model technically less complex than other anisotropic metrics. It represents a homogeneous and anisotropic geometry. One may question the relevance of anisotropic models of the observational point of view since the universe today seems to be homogeneous and isotropic. However, there is no reason, {\it a priori}, that the universe has always been so. In early times, isotropization processes of quantum nature or even thermodynamics nature \cite{Sinha} must have occurred on an essentially less regular space-time. In this sense, the study of anisotropic models in the early universe where quantum effects have great relevance is justified.

The Kantowski-Sachs space-time also has some interesting features. First, its classical solutions are known in different contexts, as well as their quantum solutions \cite{Kantowski-Sachs,Conradi,SimeoneKS,NelsonKS}. Furthermore, Kantowski-Sachs metric may coincide with the vacuum Schwarzschild solution inside the singularity. As it may also describe the interior of black holes, it can also be used to discuss their quantum states and its entropy. Models such as the Kantowski-Sachs are still interesting in order to study the behavior of the added degrees of freedom in quantum cosmological models. Finally, it is possible that the construction of a Kantowski-Sachs quantum cosmological model suggest adaptations and modifications in the quantization methods applied to cosmology. For example, in \cite{Juliobianchi}, the quantization in minisuperspace of the Bianchi I model reveals a startling discrepancy of interpretations of Bohm-de Broglie and many worlds of quantum mechanics: the quantum model is non-unitary and the norm of the wave function becomes dependent of time. Perhaps the analysis of other anisotropic quantum models is likely to explain this anomaly. In general, the condition that an operator is self-adjoint is sufficient to provide a spectral behavior, as unitarity, appropriate to a quantum system. The appearance of anomalies in functions associated with a self-adjoint operator must indicate that there are other relevant mathematical aspects that must be taken into account in the construction of a quantum cosmological model and that go beyond spectral analysis. In particular, there is suspicion that even the build procedure of the wave packet may introduce such anomalies.

Here we apply the previous investigations of quantum perfect fluid models \cite{Juliobianchi,Got,Lap,Alv,Aca,Alv2} to quantize the Kantowski-Sachs model. The Wheeler-DeWitt equation is solved for the minisuperspace and the wave function is obtained. In this procedure the time is phenomenologically introduced through one of the degrees of freedom of a stiff matter fluid. 

This article is organized as follows: In the next section, we show the Kantowski-Sachs metric and calculate the Hamiltonian function from the gravitational part. Then, the quantum cosmological model is constructed using the approach of the phenomenological time. In section \ref{sec:conclusion} we proceed a detailed discussion and conclusion of the results obtained. 

\section{The cosmological model: the choice of time and the Wheeler-DeWitt quantization}   
\label{sec:GPequation}                                

In this model we consider a flat, homogeneous and anisotropic universe described by the Kantowski-Sachs metric. According to \cite{Callan} the Kantowski-Sachs spacetime is not a spacetime itself, but only part of it. This restriction holds for the vacuum model, but disappears if a perfect fluid is coupled to gravity. Here, we introduce a perfect fluid, described by the Schutz's canonical formalism \cite{Schutz2,Schutz1} to obtain an explicit time variable related with the fluid degrees of freedom, solving two problems: the first one, described in \cite{Callan}, and the problem of the time variable that is absent in the quantum cosmological models. 

\subsection{Kantowski-Sachs spacetime}
\label{KSST}
We start with the Einstein-Hilbert action plus a boundary term
\begin{align}
S_g =\int_M d^{4}x\sqrt{-g}R+2\int_{\partial M} d^{3}x\sqrt{h}K\quad,\label{fk1}
\end{align}
where $R$ is the scalar curvature, taken as a function of $g^{\mu\nu}$ and its derivatives, $K = h_{ab}K^{ab}$ such that $K_{ab}$ is the extrinsic curvature and $h_{ab}$ is the induced metric over the three-dimensional spatial hypersurface, which is the boundary $\partial M$ of the four dimensional manifold $M$. The factor $16\pi G$ is made equal to one in the first term of equation (\ref{fk1}), where $G$ is the Newtonian gravitational constant.
\par
We will consider a model of a universe that is homogeneous but anisotropic. Thus, we shall deal with a spacetime such that the line element can be put in the more usual form \cite{Conradi}
\begin{equation}
ds^2=-N(t)^2dt^2+a(t)^2dr^2+b(t)^2({d{\theta}^2}+\sin^2\theta\,d{\varphi}^2)\quad,\label{fk2}
\end{equation}
where $a(t)$ and $b(t)$ are the scale factors and $N(t)$ denote the lapse function. This is known as Kantowski-Sachs metric \cite{Kantowski-Sachs}.
\par
We can compute the Ricci scalar $R$, which for the Kantowski-Sachs metric turns out to be
\begin{equation}
R=2\frac{\ddot{a}}{N^2a}+\frac{4}{N^2}\frac{\ddot{b}}{b}-2\frac{\dot{N}}{N^3}{\dot{a}}{a}-
4\frac{\dot{N}}{N^3}\frac{\dot{b}}{b}+\frac{4}{N^2}\frac{\dot{a}}{a}\frac{\dot{b}}{b}+
\frac{2}{N^2}\frac{\dot{b}^2}{b^2}+\frac{2}{N^2}\quad.\label{fk3}
\end{equation}
The extrinsic curvature $K$ is given by
\begin{equation}
K=-\frac{1}{N}\left(\frac{\dot{a}}{a}+2\frac{\dot{b}}{b}\right)\quad.\label{fk4}
\end{equation}
\par
Replacing the equations (\ref{fk3}) and (\ref{fk4}) in Einstein-Hilbert action (\ref{fk1}) we have the following expression
\begin{equation}
S_g=\int
dt\left(2Na-\frac{4{\dot{a}}b{\dot{b}}}{N}-\frac{2a{\dot{b}^2}}{N}\right)\quad.\label{fk5}
\end{equation}
\par
There are two types of singularities in this geometry: cigarlike singularities, when $b\rightarrow0$ and disklike singularities, when $a\rightarrow0$ \cite{Conradi}. However, the curvature of the hypersurface $^3R=\frac{2}{b^2}$ is divergent for the cigarlike singularity. Due to the absence of matter, the disklike singularity is a singularity where the right handed coordinate system becomes a left handed system. Moreover, it is an indication for the incompleteness of the vacuum Kantowski-Sachs spacetime \cite{Conradi,Callan}. The model may be completed by adding matter, e.g., in the form of stiff matter, like in this work.

With the gravitational action (\ref{fk5}), the gravitational Lagrangian of the Kantowski-Sachs model is given as follows
\begin{displaymath}
L_g=-\frac{{\dot{a}}b{\dot{b}}}{N}-\frac{a{\dot{b}^2}}{2N}+\frac{Na}{2}\quad.
\end{displaymath}

In order to deal with a mathematically tractable problem \cite{Conradi}, let us perform the reparametrization
\begin{displaymath}
a\rightarrow{a};\;\;\;\;b\rightarrow{\frac{c}{a}};\;\;\;\;\dot{b}=\frac{\dot{c}}{a}-\frac{\dot{a}c}{a^2}\quad.
\end{displaymath}
Thus, the purely gravitational Lagrangian is written as
\begin{equation}
L_g=\frac{Na}{2}+\frac{\dot{a}^2c^2}{2Na^3}-\frac{\dot{c}^2}{2Na}\quad.\label{fk6}
\end{equation}
Using the canonical formalism the gravitational Hamiltonian can be expressed as
\begin{equation}
H_g=\frac{a}{2c^2}[a^2p_a^2-c^2p_c^2-c^2]\;.\label{HG1}
\end{equation}

In the next section we will introduce the matter as a barotropic perfect fluid described by the thermodynamic variables.

\subsection{Phenomenological time: Schutz's canonical formalism}
\label{SF}

Schutz \cite{Schutz2,Schutz1} introduced, in the context of general relativity, a representation for a perfect fluid in terms of velocity potentials. The formalism describes the dynamics of a relativistic fluid in interaction with the gravitational field.

The hydrodynamic equations are put in Eulerian form, with the four-velocity  expressed in terms of five potentials $\phi, \zeta, \beta, \theta$ and $S$: 

\begin{equation}
U_{\nu} = \frac{1}{\mu}(\phi_{,\nu} + \zeta\beta_{,\nu} + \theta S_{,\nu}) \label{potencial}\,\, ,
\end{equation}

\noindent each satisfying its own equation of motion. In this expression  $\mu$ is the specific enthalpy, the variable $S$ is the specific entropy, while the potentials $\zeta$ and $\beta$ are connected with rotation and are absent in this model. The variables $\phi$ and $\theta$ have no clear physical meaning. 

The derived dynamical equations  provide a description of the usual equivalent hydrodynamic equations based on the divergence of the energy-momentum tensor. From this, follows an especially simple action,

\begin{equation}
S_{f}=\int\,d^4x\sqrt{-g}\,p \,\, , \label{fk7} 
\end{equation}

\noindent where $p$ is the pressure, which is associated to the energy density by the equation of state $p = \alpha\rho$.

It is important to mention that the Einstein's equations for a perfect fluid can be found when we vary the action (\ref{fk7}) plus the Einstein-Hilbert action in respect to the metric tensor. On the another hand, the changes in relation to the velocity potentials is similar to the Eulerian equations of evolution of the fluid,

\begin{displaymath}
U^\nu\,{\phi}_{,\nu}=-\mu \, ; \, \quad U^\nu\,{\zeta}_{,\nu}=0 \,\, ; \, \quad U^\nu\,{\beta}_{,\nu}=0\, ;
\end{displaymath}

\begin{equation}
U^\nu\,{\theta}_{,\nu}=T_s\,  ; \, \quad U^\nu\,{S}_{,\nu}=0 \,\, ; \, \quad ({\rho_{0}}U^\nu)_{;\nu}=0 \label{movimentos}\, .
\end{equation}
 
\noindent Here $T_s$ is the temperature and $\rho_{0}$ is the rest mass density of the fluid.

The basic thermodynamic relations take the form

\begin{equation}
\rho=\rho_0(1+\Pi)\, ;\, \quad \mu=(1+\Pi) + \frac{p}{\rho_{0}} \, ; \, \quad T_sdS=d\Pi +p d\bigg(\frac{1}{\rho_{0}}\bigg) \, ,  \,
\end{equation}

\noindent $\Pi$ being the specific internal energy.

If we write 

\begin{equation}
d\Pi +p d\bigg(\frac{1}{\rho_{0}}\bigg)=(1+\Pi)d[ln(1+\Pi)-\alpha\,ln\rho_0]\,\, ,
\end{equation}

\noindent we can identify $T_s=1+\Pi\,\, \mbox{and} \,\, S=ln(1+\Pi)/{\rho_{0}}^{\alpha}$. 

After a few mathematical steps, the energy density and pressure reduces to

\begin{equation}
\rho=\left(\frac{\mu}{\alpha+1}\right)^{1+\frac{1}{\alpha}}e^{-\frac{S}{\alpha}}\,\,\, \mbox{and}\,\,\, p=\alpha\left(\frac{\mu}{\alpha+1}\right)^{1+\frac{1}{\alpha}}e^{-\frac{S}{\alpha}}\,\, \label{pressure}.
\end{equation}

By means of the normalization condition

\begin{equation}
U^{\nu}U_{\nu}=-1\, \, ,\label{131}
\end{equation}

\noindent once can express $\mu$ in terms of the potentials

\begin{equation}
\mu=\frac{1}{N}(\dot{\phi}+\theta\dot{S})\,\,\label{entalpia2} .
\end{equation}

Finally, by using (\ref{pressure}) an (\ref{entalpia2}) in the matter action (\ref{fk7}), it is possible to obtain the Lagrangian of the fluid:

\begin{equation}
L_f=\frac{c^2}{a}\,\,N^{-1/{\alpha}}\,\, \frac{\alpha}{({\alpha}+1)^{1+1/{\alpha}}}\,\,(\dot{\phi}+
{\theta}\dot{S})^{1+1/{\alpha}}\,\, e^{-S/{\alpha}} \quad . \label{fk9}
\end{equation}

The fluid conjugated momenta are derived from the above Lagrangian, written in terms of the canonical variables
\begin{eqnarray}
p_{\phi}&=&\frac{c^2}{a}\,\,N^{-1/{\alpha}}\,\, \frac{1}{({\alpha}+1)^{1/{\alpha}}} \,\,(\dot{\phi}+
{\theta}\dot{S})^{1/{\alpha}}\,\,  e^{-S/{\alpha}}\quad,\nonumber\\
p_S&=&\theta p_{\phi}\quad,\nonumber\\
p_{\theta}&=&0\quad ,
\end{eqnarray}
and again by the canonical formalism we obtain

\begin{equation}
H_f=\left(\frac{a}{c^2}\right)^{\alpha}e^S \,p_{\phi}^{\alpha+1}\;.\label{Hf1}
\end{equation}

This can be put in a more suggestive form by means of the canonical transformations
\begin{equation}
T=p_S e^{-S}{p_{\phi}}^{-(\alpha+1)}\, , \,  \quad p_{T} = {p_{\phi}}^{(\alpha + 1)} e^{S}\, ; \, \quad \overline{\phi}=\phi -(\alpha+1)\frac{p_{S}}{p_{\phi}}\, , \, \quad 
\overline{p}_{\phi}=p_{\phi}\, , \label{tc1}
\end{equation}
and the super-Hamiltonian (\ref{Hf1}) can be written in the final form: 
\begin{equation}
H_f=\left(\frac{a}{c^2}\right)^{\alpha}\, p_T\;.\label{Hf2}
\end{equation}

Time introduced via Schutz's formalism is a global phase time. In fact, to be a global phase time the function must be monotonously increasing, defined in phase space and satisfying the relation
\begin{equation}
[t, H] > 0\quad.
\end{equation}
In particular, the time introduced in equations (\ref{tc1}) and (\ref{Hf2}) is an extrinsic time. If the function defined in phase space is of type $t(q_i,p_i)$, it is called extrinsic time. But, if the function has the form $t(q_i)$ it is called intrinsic time \cite{Simeonelivro}. The idea of a global phase time in anisotropic universes of this type is well exposed in reference \cite{SimeoneKS}. It is possible, in some contexts, to connect the time with the geometric origin of the quantum theory. In the Kantowski-Sachs universe the time can be a global phase factor if, throughout some gauge transformations, we can make it obeys the relation $[t,H]>0$. 
 
\subsection{Kantowski-Sachs quantum cosmology with perfect fluid}   
\label{sec:gPequation}                                

In this section, we describe the Schutz's formalism in Kantowski-Sachs spacetime. The total Hamiltonian $H=H_g+H_f$ is given by the gravitational part $H_g$ plus the fluid part $H_f$, that is
\begin{equation}
H=\frac{a}{2c^2}\left[a^2p_a^2-c^2p_c^2-c^2+2\left(\frac{a}{c^2}\right)^{\alpha-1}p_T\right]\quad.\label{fk10}
\end{equation}
The conjugate momentum $p_T$ associated to matter appears linearly in $H$, and in this way a Schr\"odinger equation can be obtained with the matter variable playing the role of time. Therefore, all tools of ordinary quantum mechanics can, in principle, be employed in order to obtain predictions regarding the evolution of the universe.

As usual, the Kantowski-Sachs model is quantized following Dirac's quantization scheme by turning the variables $p_a$, $p_c$ and $p_T$ into operators, which fulfill the standard commutation relations
\begin{displaymath}
p_a\rightarrow\hat{p_a}=-i \frac{\partial}{\partial a}\,; \quad p_c\rightarrow\hat{p_c}=-i \frac{\partial}{\partial c}\,; \quad p_{T}\rightarrow\hat{p_{T}}= -i \frac{\partial}{\partial T}\quad.
\end{displaymath}

In quantum cosmology, the Hamiltonian formulation of general relativity is employed through the ADM decomposition of the geometry \cite{ADM}, and a Schr\"odinger-like equation, the Wheeler-DeWitt equation in the field representation, is constructed, which determines the wave function of the universe as a whole. Performing in this way, the Hamiltonian (\ref{fk10}) becomes an operator, which annihilates the physical states: $H = 0 \rightarrow \hat H\Psi = 0$. Using the new time reparametrization $t\rightarrow2T$ we have the following Wheeler-DeWitt equation
\begin{equation}
\left(-a^2\frac{\partial^2}{\partial a^2}+c^2\frac{\partial^2}{\partial c^2}-c^2+i\left(\frac{a}{c^2}\right)^{\alpha-1}\frac{\partial}{\partial t}\right)\Psi(a,c,t)=0\quad.\label{Hprofk}
\end{equation}

Functions in the classical phase space of this theory can commute. However, the associated operators defined above in Hilbert space, in general, do not commute. As a result, the order in which the product of two operators is written becomes significant, which leads to an ordering problem. The most direct way to write the operator is one that makes it symmetrical, which is a precondition for the operator to be self-adjoint,  which will make it a physical observable. A short discussion about the self-adjointness of the operator is made in reference \cite{nivaldolivro}. Thus, we have
\begin{equation}
(\hat{H}\psi,\phi)=(\psi,\hat{H}\phi)\quad.\label{fk12}
\end{equation}
We propose the following ordering
\begin{equation}
a^2\partial_a^2\rightarrow
a^2\partial_a^2+q_a\partial_a=X\quad.\label{fk13}
\end{equation}
The inner product $(\psi,X\phi)$ is, by definition, written as
\begin{equation}
(\psi,X\phi)=\int\psi^*\left(a^2\partial_a^2+q_a\partial_a\right)\phi~da\quad.\label{k14}
\end{equation}
After some integrations by parts, we can write
\begin{equation}
(\psi,X\phi)=\int\psi^*\left(a^2\partial_a^2+2a\partial_a\right)\phi~da=(X\psi,\phi)\quad,\label{k15}
\end{equation}
and similarly, if
\begin{equation}
c^2\partial_c^2\rightarrow
c^2\partial_c^2+q_c\partial_c=Y\quad,\label{k13}
\end{equation}
we will have
\begin{equation}
(\psi,Y\phi)=\int\psi^*\left(c^2\partial_c^2+2c\partial_c\right)\phi~dc=(Y\psi,\phi)\quad,\label{k16}
\end{equation}
where $q_a = 2a$, in equation (\ref{k15}), and $q_c = 2c$, in equation (\ref{k16}), to ensure the hermiticity of $\hat{H}$ and the positivity of the Schr\"odinger inner product. A more detailed discussion can be found in \cite{SimeoneKS}. With this new ordering, the Wheeler-DeWitt equation can now be rewritten as
\begin{equation}
\left[-a^2\frac{\partial^2}{\partial{a^2}}-2a\frac{\partial}{\partial{a}}+c^2\frac{\partial^2}{\partial{c^2}}+
2c\frac{\partial}{\partial{c}}-c^2+
i\left(\frac{a}{c^2}\right)^{\alpha-1}\frac{\partial}{\partial{t}}\right]\Psi(a,c,t)=0\quad.\label{k17}
\end{equation}
Note that the signature of the kinetic term is hyperbolic, something that often occurs in anisotropic models. On the assumption of stationary solutions,
$\Psi(a,c,t)=\psi(a,c)\,e^{-iEt}$, we are led to
\begin{equation}
\left[c^2\frac{\partial^2}{\partial{c^2}}+2c\frac{\partial}{\partial{c}}-c^2-a^2\frac{\partial^2}{\partial{a^2}}-
2a\frac{\partial}{\partial{a}}\right]\psi(a,c)=E\left(\frac{a}{c^2}\right)^{\alpha-1}\psi(a,c)\quad.\label{k19}
\end{equation}

Analytical solutions of this equation are possible only for the particular case where $\alpha = 1$ (stiff matter). In addition to the reasons discussed above, this case is reasonable to check the connection between Schutz's formalism and cosmological models with anisotropic metrics. Thus, with $\alpha=1$, we have
\begin{equation}
\left(c^2\frac{\partial^2}{\partial{c^2}}+2c\frac{\partial}{\partial{c}}-c^2-a^2\frac{\partial^2}{\partial{a^2}}-2a\frac{\partial}{\partial{a}}-
E\right)\psi(a,c)=0\quad.\label{k20}
\end{equation}
The operators $a^2{\partial^2_{a}} + 2a\partial_{a}$ and $c^2{\partial^2_{c}} + 2c\partial_{c}$ in $L^{2}(0,\infty)$ are unitarily equivalent to the operators $\partial^2_a$ and $\partial^2_c$ in $L^{2}(-\infty,\infty)$ \cite{nivaldo}, which are self-adjoint (in fact, if two operators $A$ and $B$ are unitarily equivalent, $B$ enjoy all the spectral properties that characterize $A$). This mapping is defined by the transformation
\begin{displaymath}
\tilde{\psi}(x,y)=e^{-x/2 -y/2}\psi(e^{-x},e^{-y})\quad,
\end{displaymath}
where $a=e^{-x}$ and $c=e^{-y}$. Therefore, the equation (\ref{k20}) is modified to
\begin{equation}
\left(\frac{\partial^2}{\partial{y^2}}-\frac{\partial^2}{\partial{x^2}}- e^{-2y}\right)\tilde{\psi}(x,y) = E \tilde{\psi}(x,y)\quad,\label{fk15}
\end{equation}
where we assume that $E > 0$.

Applying the method of separation of variables, we obtain
\begin{equation}
\tilde{\psi}(x,y)=X(x)Y(y)\quad,\label{fk16}
\end{equation}
that leads to the equations
\begin{equation}
\frac{1}{X}\frac{\partial^2 X}{\partial x^2} = -{\lambda}^2\quad,\label{fk17}
\end{equation}
where $-\lambda^2$ is a separation constant and,
\begin{equation}
\frac{1}{Y}\frac{\partial^2 Y}{\partial y^2} - ({\sigma}^2 + e^{-2y}) = 0 \quad,\label{fk18}
\end{equation}
where $\sigma^2=E - {\lambda^2}$.

The general solution to the wave function of the universe follows from the above equations

\begin{equation}
\Psi_{\lambda\sigma}(x,y,t)= [A_1\,\sin(\lambda x) + A_2\,cos(\lambda x)] \, \, [B_1\,I_{\sigma}(e^{-y}) + B_2\, K_{\sigma}(e^{-y})]e^{-i({{\lambda}^2+ \sigma^2})t}\quad,\label{fk19}
\end{equation}
where $A_1$, $A_2$, $B_1$ and $B_2$ are integration constants, $I_{\sigma}$ and $K_{\sigma}$ are modified Bessel functions and we assumed that $\sigma\in \mathbb{C}$ and $\lambda\in \mathbb{R}$. 
\par
Having this general solution, we need to investigate under what conditions the above wave function satisfies the boundary conditions, i.e., $\Psi(a, c, t ) = 0$ (when $a\rightarrow{0}$ and $c\rightarrow{0}$). Under this restriction, we obtain the following wave function
\begin{equation}
\Psi_{\lambda\sigma}(x, y, t)= K_\sigma(e^{-y})e^{\pm i \lambda x}e^{-i(\lambda^2+\sigma^2)t}\quad.\label{fk20}
\end{equation}

With the boundary conditions, the wave packets may be constructed by taking superpositions of the $\lambda$ and $\sigma$. The general structure of these superpositions is
\begin{equation}
\Psi(x,y,t)=\int^{\infty}_{0}\int^{\infty}_{0}C(\lambda,\sigma)e^{- i \lambda x} \,  K_\sigma(e^{-y})e^{-i{\lambda^2}t}e^{-i{\sigma^2}t}d\lambda\,d\sigma\quad. \label{fk21}
\end{equation}

 Note that the above solution admits the following factorization
 
	\begin{equation}
	   \Psi(x,y,t)=\mathcal{I}_{\sigma}(x,t)\mathcal{I}_{\lambda}(y,t) \,, \label{fator}
	\end{equation}
\noindent	where
	\begin{eqnarray}
	   \mathcal{I}_{\lambda}&=&\int^{\infty}_{0} C_{1}(\lambda)e^{-i({\lambda^2}t + \lambda x)}\,d\lambda  \quad,\label{fk22} \\
     \mathcal{I}_{\sigma}&=&\int^{\infty}_{0} C_{2}(\sigma) K_{\sigma}(e^{-y})\, e^{-i\sigma^2 t}\, d\sigma\quad .\label{fk25}
	\end{eqnarray}
	
	For integral (\ref{fk22}), if we adopt  $C_1(\lambda)= e^{-i\lambda^2 t_0}$ and $\lambda^2t+\lambda x=(\lambda \sqrt{t+t_0}+\frac{x}{2\sqrt{t+t_0}})^2-\frac{x^2}{4(t+t_0)}$, we obtain 
	\begin{equation}
	\label{flavio1}
	   \mathcal{I}_{\lambda}=e^{-i\frac{x^2}{4(t+t_0)}}\int^{\infty}_{0}\left\{\cos\left[\left(\lambda \sqrt{t+t_0}+\frac{x}{2\sqrt{t+t_0}}\right)^2\right]-i\sin\left[\left(\lambda \sqrt{t+t_0}+\frac{x}{2\sqrt{t+t_0}}\right)^2\right]\right\} \,d\lambda \,,
	\end{equation}
	where $t_0$ is an arbitrary constant. Introducing the variable $u = \lambda \sqrt{t+t_0}+\frac{x}{2\sqrt{t+t_0}}$, the integral (\ref{flavio1}) takes the form
  \begin{equation}
	   \mathcal{I}_{\lambda}={\frac{e^{-i\frac{x^2}{4(t+t_0)}}}{\sqrt{t+t_0}}\int^{\infty}_{\frac{x}{2\sqrt{t+t_0}}}\left(\cos{u^2}-i\sin{u^2} \right)}\,du \, ,
	\end{equation}
	\noindent which has the solution given by
	\begin{equation}
	   \mathcal{I}_{\lambda}=\frac{e^{-i\frac{x^2}{4(t+t_0)}}}{\sqrt{t+t_0}}\sqrt{\frac{\pi}{8}}\left\{\left[1-2C\left(\sqrt{\frac{2}{\pi}}\frac{x}{2\sqrt{t+t_0}}\right)\right]-i\left[1-2S\left(\sqrt{\frac{2}{\pi}}\frac{x}{2\sqrt{t+t_0}}\right)\right]\right\} \, ,
	\end{equation}
	where $C$ e $S$ are the Fresnel integrals
	\begin{eqnarray}
	   C(z)=\int_{0}^{z}{\cos\left(\frac{\pi}{2}\theta^2\right)}d\theta \,, \\
		 S(z)=\int_{0}^{z}{\sin\left(\frac{\pi}{2}\theta^2\right)}d\theta \,.
	\end{eqnarray}
	
	For the probability density, we have
	\begin{eqnarray}
	   \label{eq:I_lambda}
	   \left|\mathcal{I}_{\lambda}\right|^2&=&\frac{\pi}{4(t+t_0)}\bigg{\{}1+2\bigg{[}C\left(\frac{1}{\sqrt{2\pi}}\frac{x}{\sqrt{t+t_0}}\right)\left(C\left(\frac{1}{\sqrt{2\pi}}\frac{x}{\sqrt{t+t_0}}\right)-1\right) + \\ \nonumber
		&+&S\left(\frac{1}{\sqrt{2\pi}}\frac{x}{\sqrt{t+t_0}}\right)\left(S\left(\frac{1}{\sqrt{2\pi}}\frac{x}{\sqrt{t+t_0}}\right)-1\right)\bigg{]}\bigg{\}} \,.
	\end{eqnarray}
	
\begin{figure}[h!]
    \centering
       \includegraphics[width=0.48\textwidth]{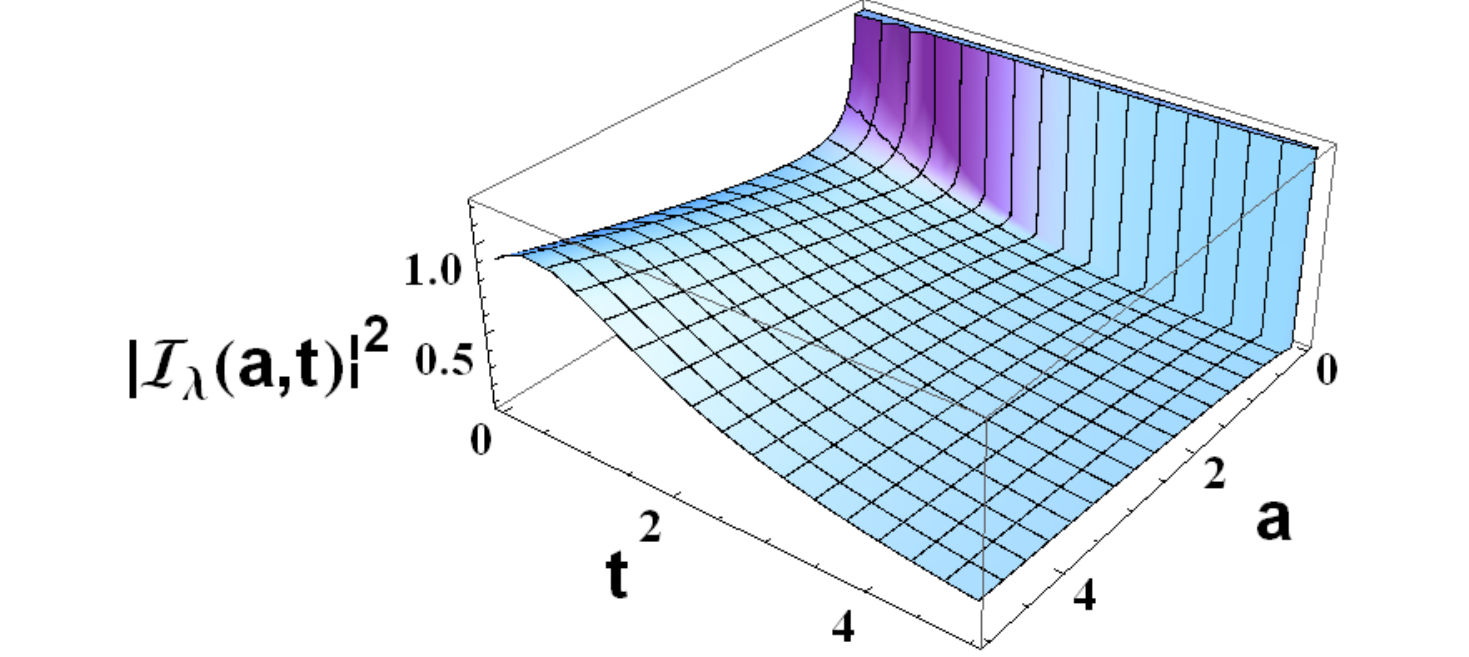}
		   \includegraphics[width=0.48\textwidth]{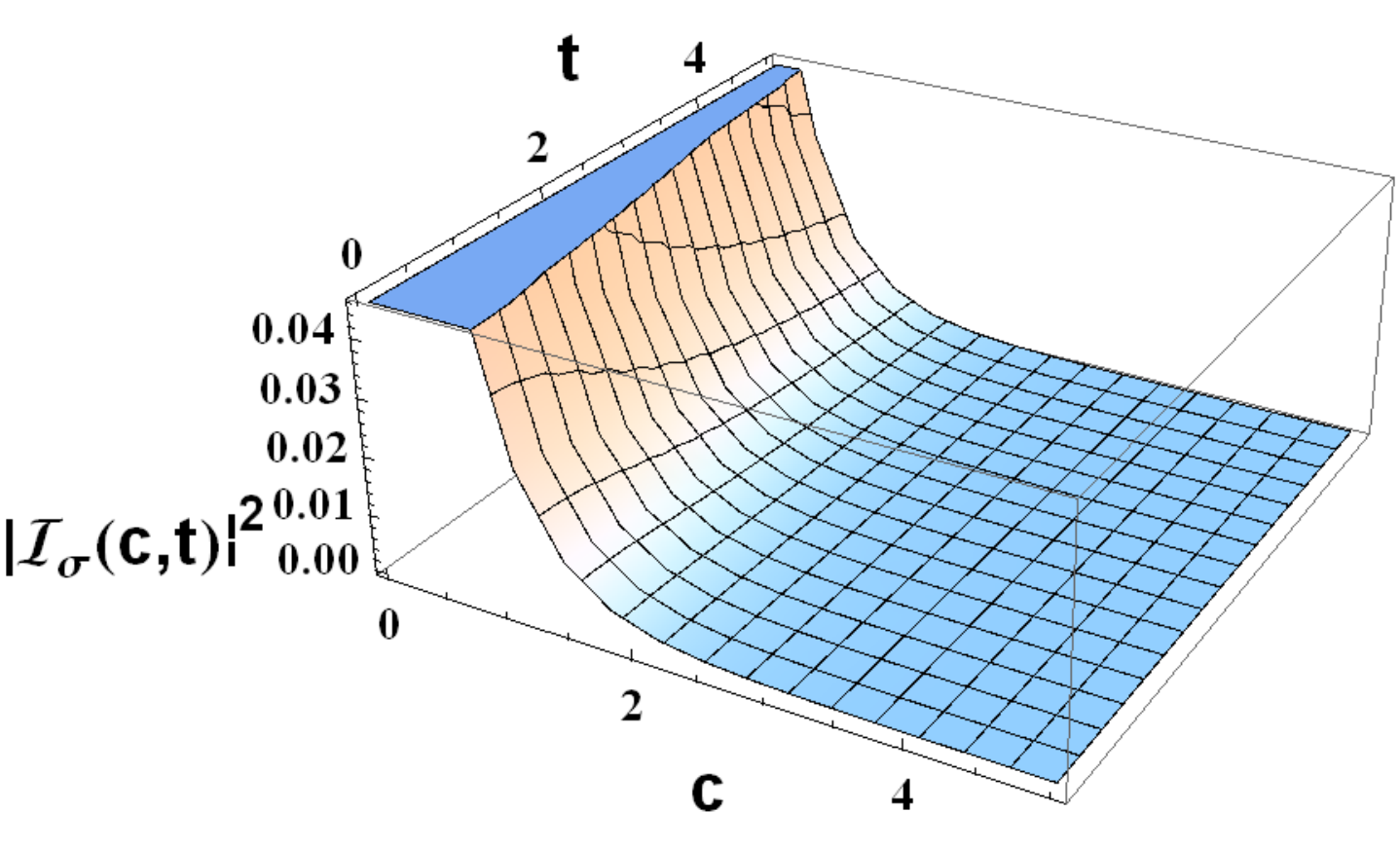}
		\caption{The behavior of $|\mathcal{I}_\lambda|^2$, which is given by equation (\ref{eq:I_lambda}), and $|\mathcal{I}_\sigma|^2$, which is found with equation (\ref{eq:I_sigma}). Here we adopt $t_0=1.5$.}
		\label{fig:I}
\end{figure}

	A similar procedure is employed to solve the integral (\ref{fk25}). If we put the Bessel function in the integral form
	\begin{equation}
	   K_{\sigma}(c)=\int_0^{\infty}{e^{-c\cosh{z}}\cosh{i\sigma z}}dz \,,
	\end{equation}
	
	\noindent and we introduce $C_2(\sigma)=e^{-i\sigma^2t_0}$ and $u_{\pm}=\sigma\sqrt{t+t_0}\pm\frac{z}{2\sqrt{t+t_0}}$, we obtain
	\begin{equation}
	   \label{eq:I_sigma}
	   \mathcal{I}_\sigma = \sqrt{\frac{\pi}{8}}\frac{(1-i)}{\sqrt{t+t_0}}\int_0^{\infty}{e^{-c\cosh{z}+i\frac{z^2}{4(\sqrt{t+t_0})}}}dz \,.
	\end{equation}
		The expressions found for $|\mathcal{I}_{\lambda}|^2$ and $|\mathcal{I}_\sigma|^2$ are plotted in figure \ref{fig:I}.
		
\begin{figure}[h!]
    \centering
       \includegraphics[width=0.48\textwidth]{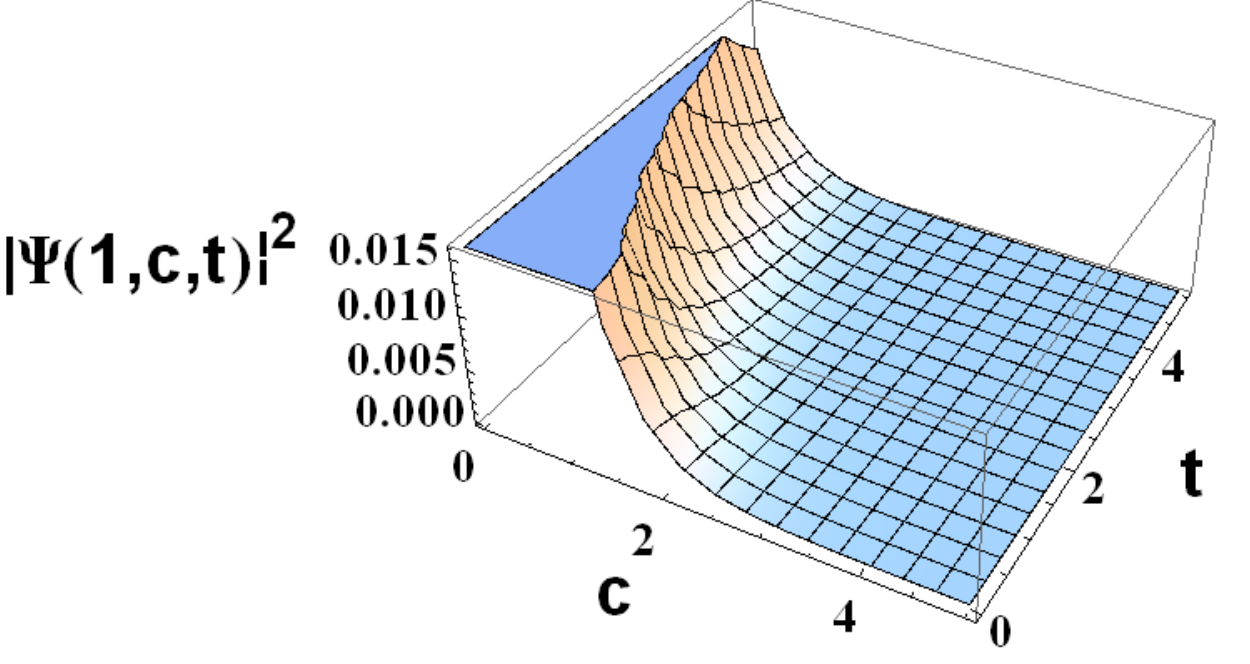}
		   \includegraphics[width=0.48\textwidth]{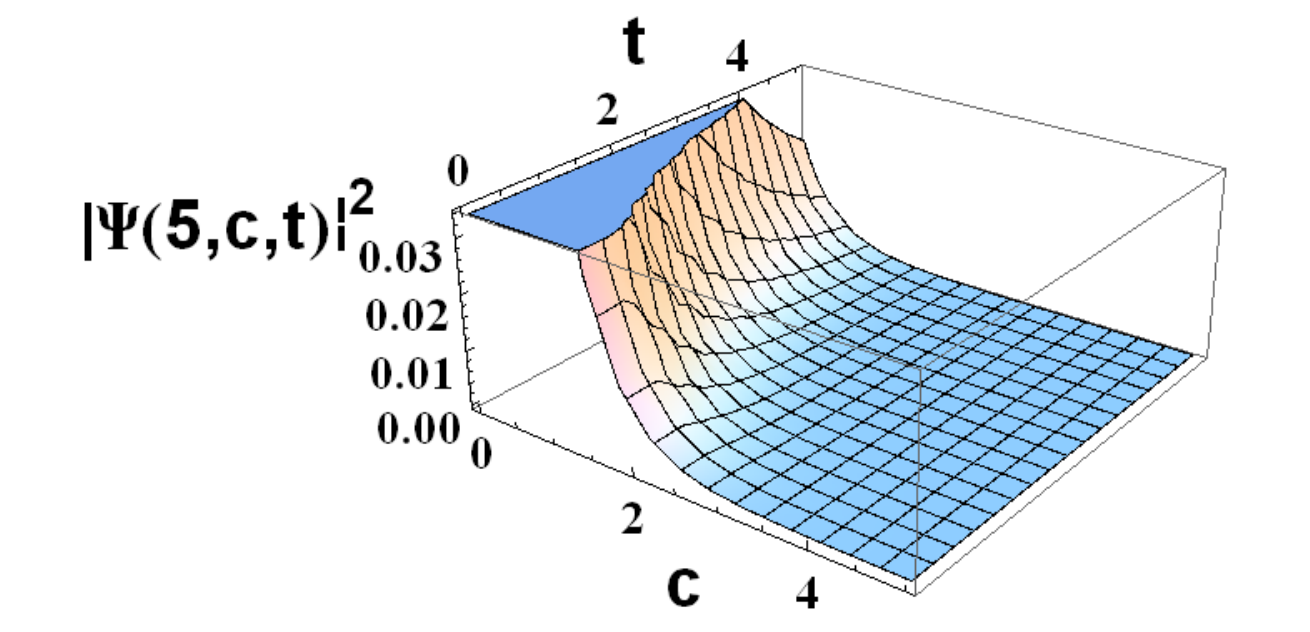} \\
			 \includegraphics[width=0.48\textwidth]{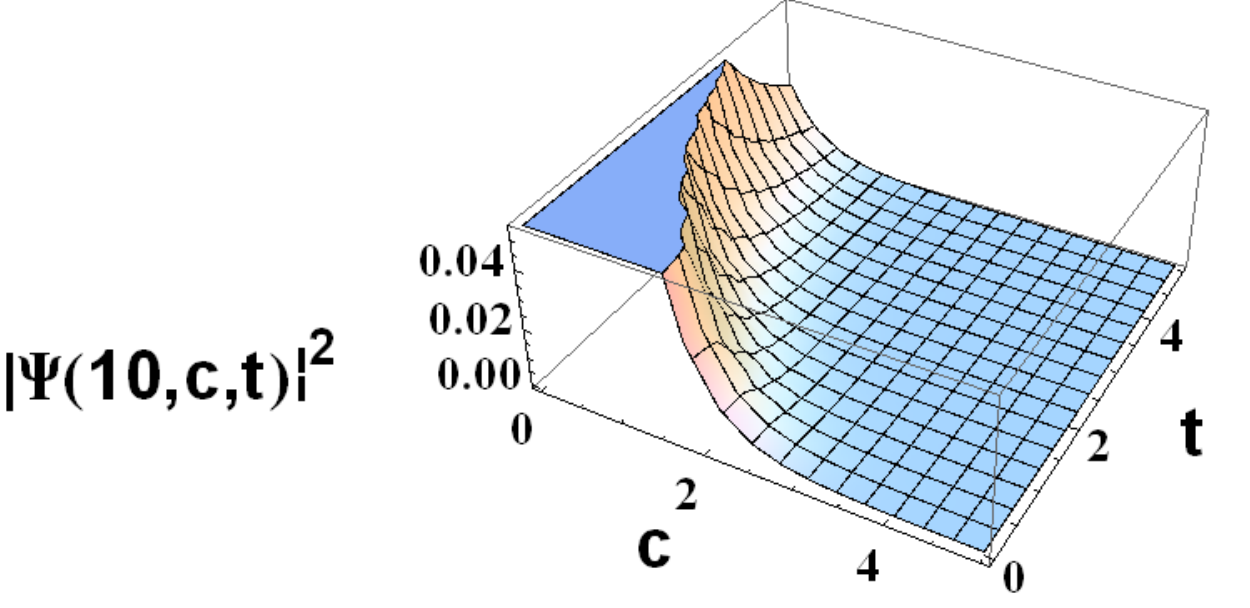}
		\caption{The wave packet as a function of $c$ and $t$, for $a=0$, $5$, $10$ and $t_0=1.5$.}
		\label{fig:pacote_ct}
\end{figure}

\begin{figure}[h!]
    \centering
       \includegraphics[width=0.48\textwidth]{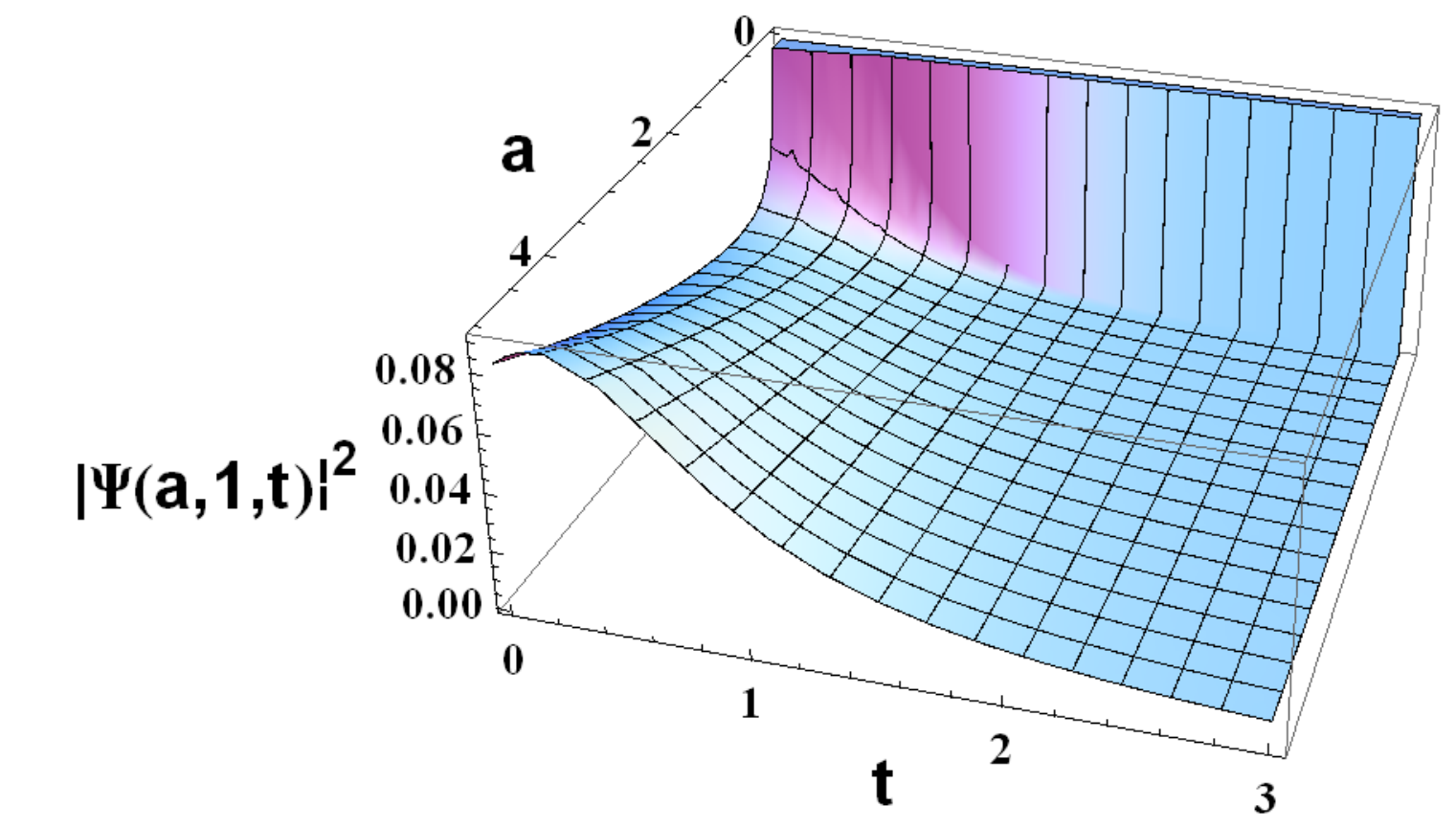}
		   \includegraphics[width=0.48\textwidth]{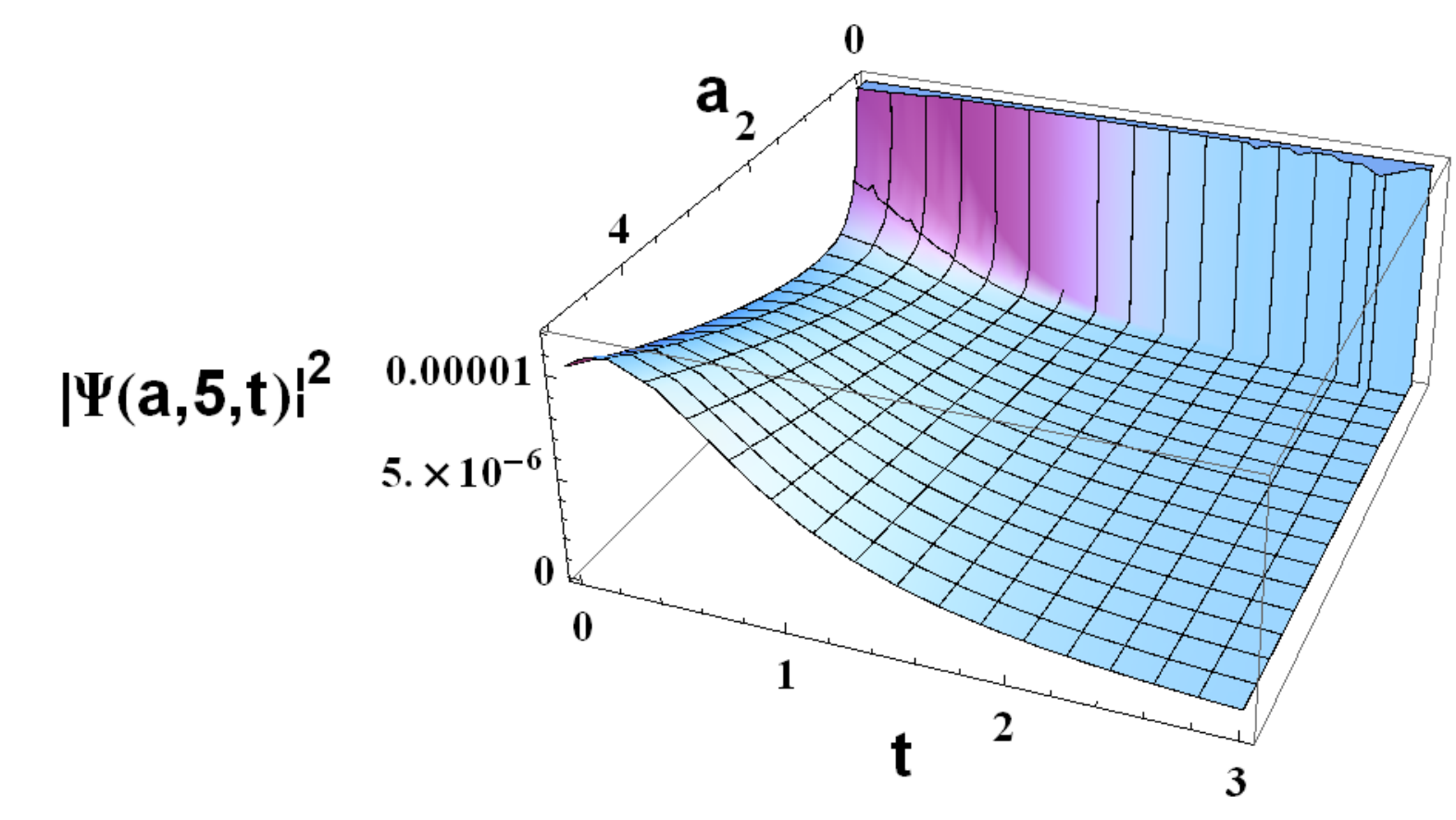} \\
			 \includegraphics[width=0.48\textwidth]{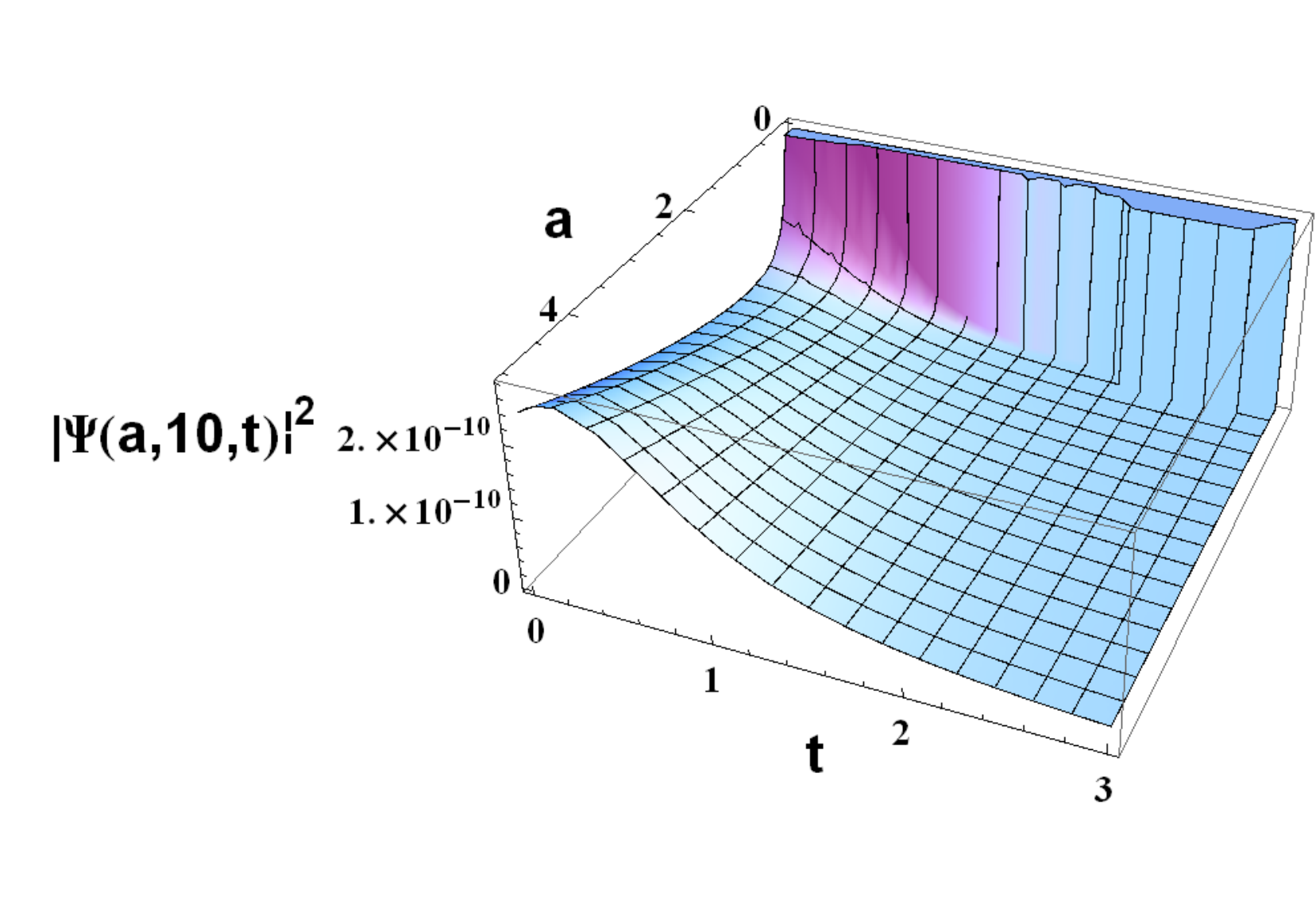}
		\caption{The wave packet as a function of $a$ and $t$, for $c=0$, $5$, $10$ and $t_0=1.5$.}
		\label{fig:pacote_at}
\end{figure}

\begin{figure}[h!]
    \centering
       \includegraphics[width=0.48\textwidth]{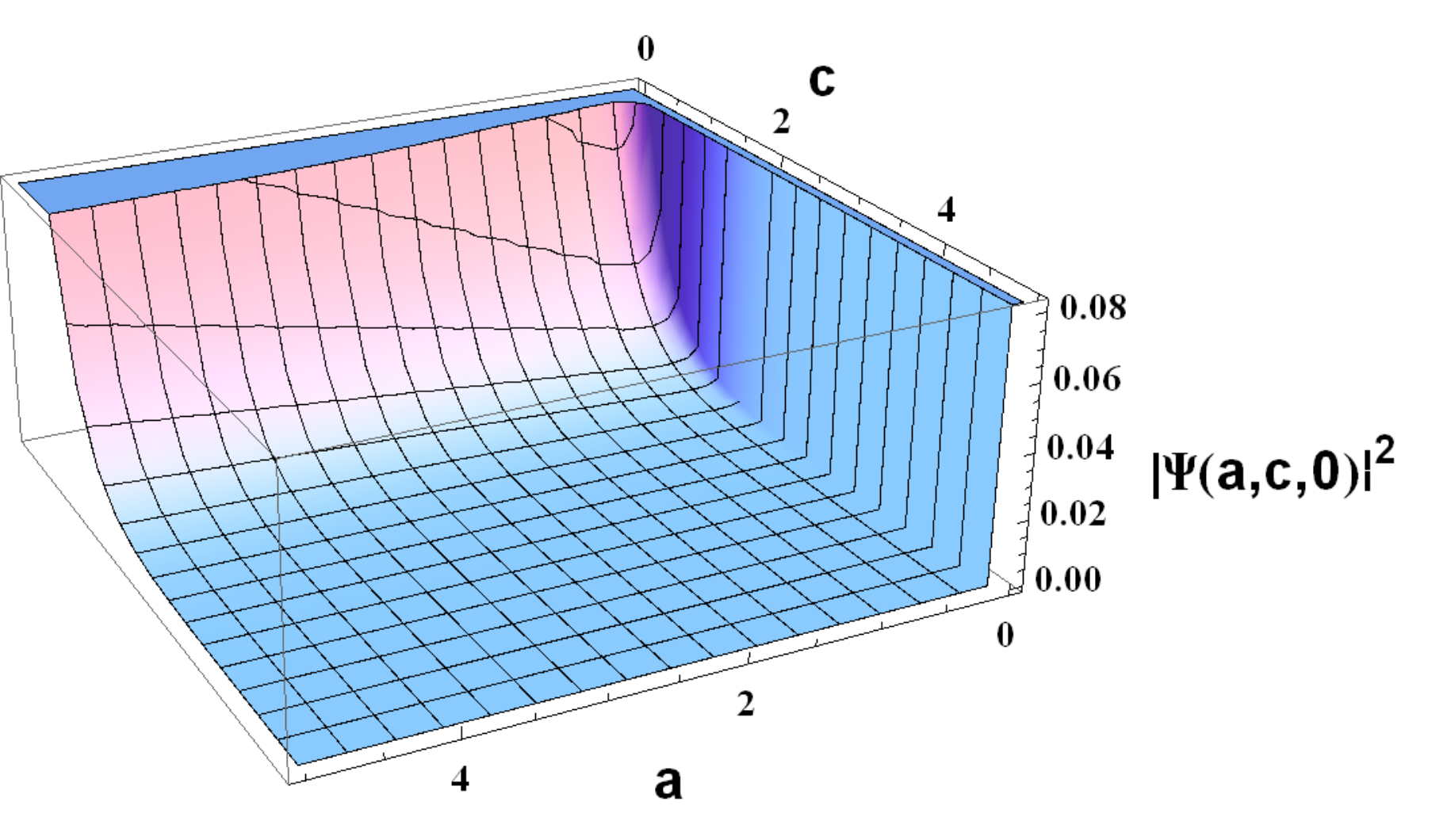}
		   \includegraphics[width=0.48\textwidth]{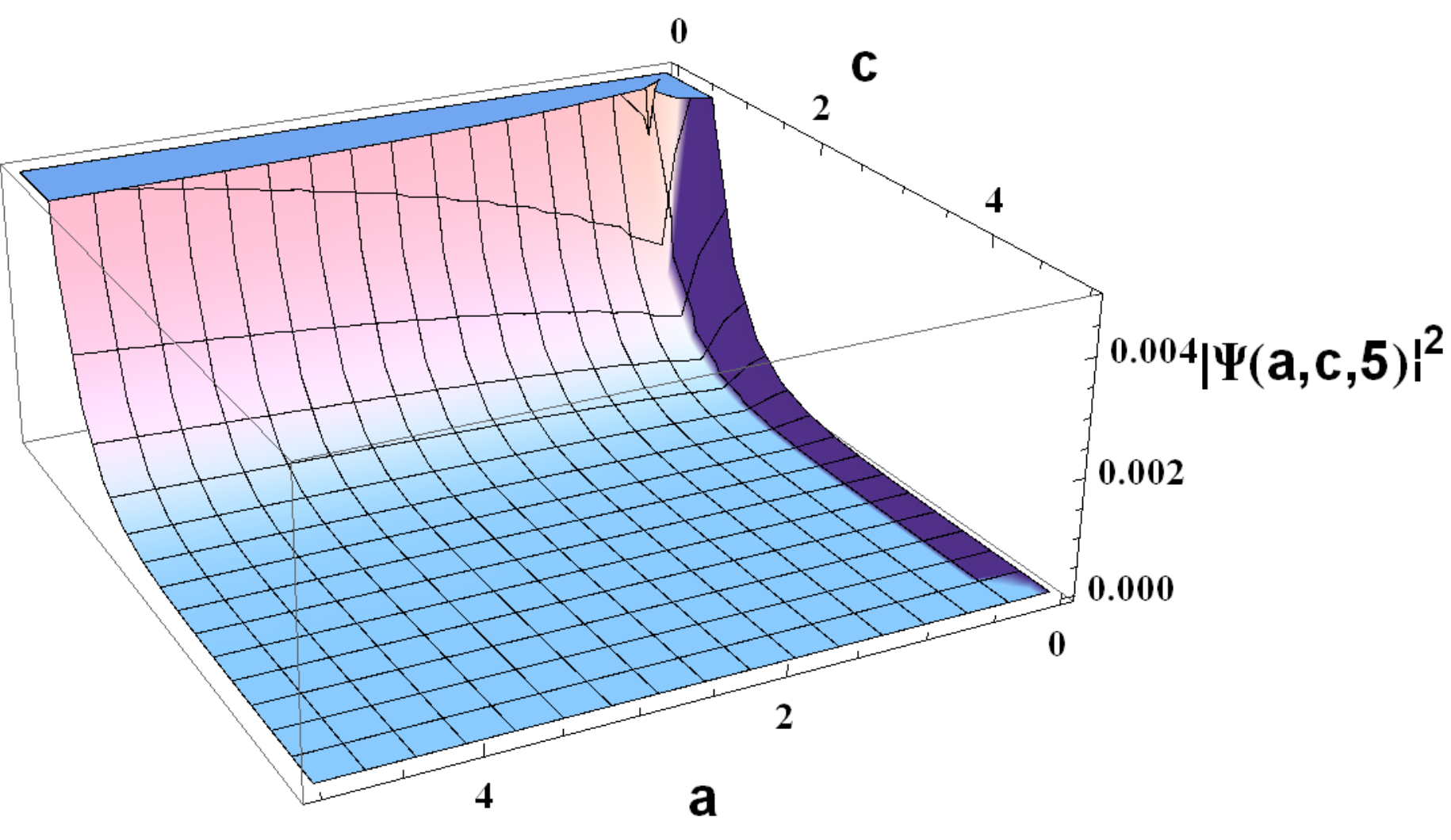} \\
			 \includegraphics[width=0.48\textwidth]{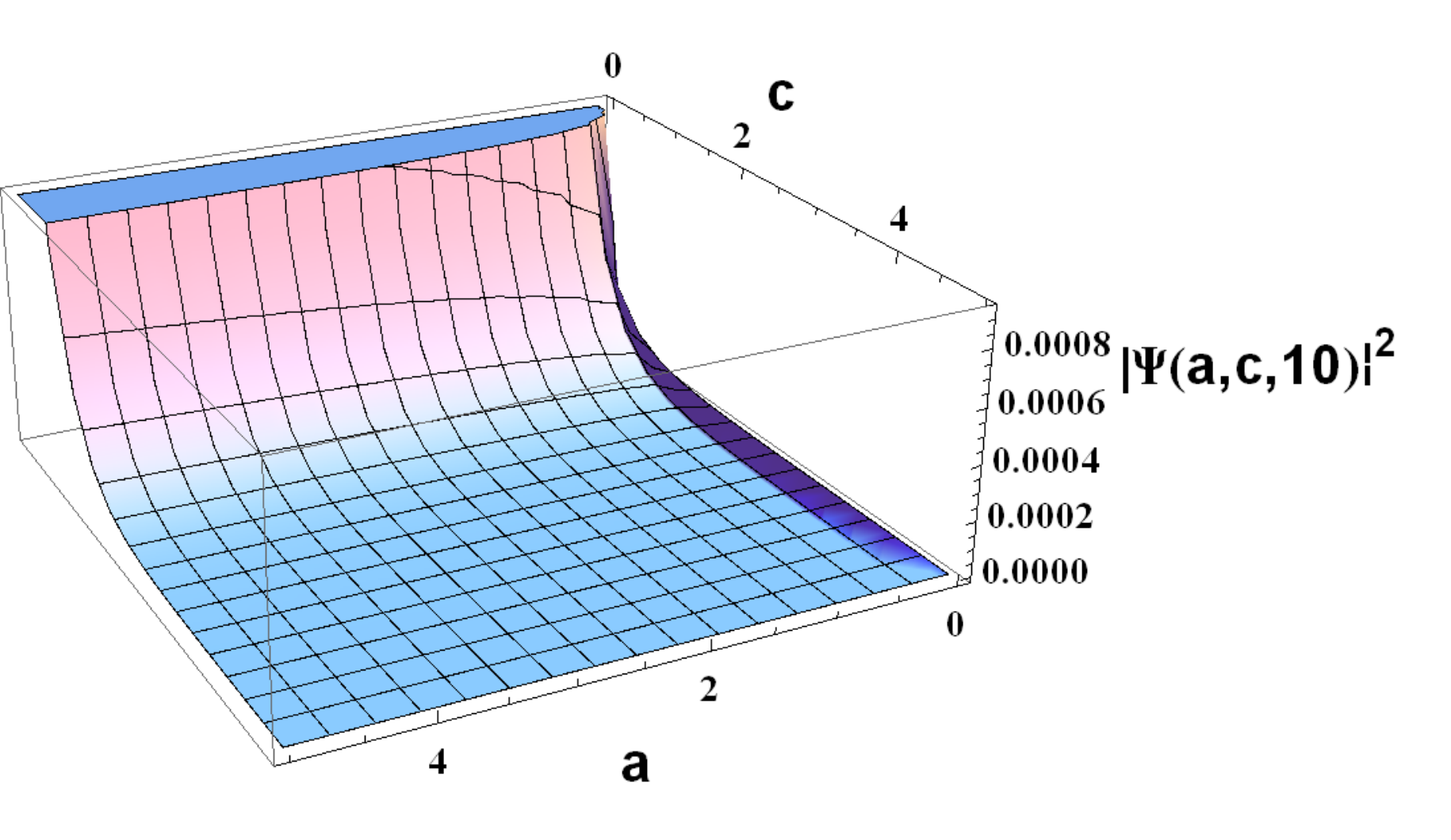}
		\caption{The wave packet as a function of $a$ and $c$, for $t=0$, $5$, $10$ and $t_0=1.5$.}
		\label{fig:pacote_ac}
\end{figure}

The relevant plots for the unnormalized probability density function is given in figures \ref{fig:pacote_ct}, \ref{fig:pacote_at} and \ref{fig:pacote_ac}. We note that the norm is time dependent and therefore the evolution is not unitary, which is also found in the context of the Bianchi model \cite{Juliobianchi}. While this result may reinforce evidence of the existence of an incompatibility between anisotropic models and Schutz formalism, it should be noted that, according to the equation (\ref{fk15}) the operator associated with the wave function obtained is self-adjoint. Since the initial conditions are also well established, the loss of unitarity must have been caused by the wave packet chosen. Although the packet is nothing more than a superposition of plane wave functions, it is possible that certain choices are not able to locate the solution in a finite region of space or to preserve the norm independent of time. This may explain the occurrence of pathologies observed in this sense in quantum cosmological models. The wave packet could have been built someway else, which in this case means other choices for $C_1(\lambda)$ and $C_2(\sigma)$. The choices made here allowed us to find an analytical expression for the wave function that although useful, may not lead to the best physical description. In our case the price to pay is a unnormalizable wave packet, since the normalization integral diverges in the $x$ direction.

\section{Conclusions and Discussions}
\label{sec:conclusion}
In this paper, we studied the Kantowski-Sachs quantum cosmological model. The introduction of a time variable in the theory was performed using the Schutz's formalism, which associates the time with degrees of freedom of a perfect fluid coupled to gravitation, chosen in this work as stiff matter.

The Schutz's formalism enables the introduction of time in theory to occur consistently, allowing the Wheeler-DeWitt equation in this minisuperspace to have an explicitly time-dependent solution. However, the model is not unitary since the norm is time-dependent, which has been observed in the context of Bianchi I model \cite{Juliobianchi}. The guarantee that the operator associated with the model is self-adjoint was not able to eliminate anomalies already detected in anisotropic cosmological models such as the loss of unitarity. This indicates that certain conditions may arise at later stages of the model construction, as in the build procedure of the wave packet. In fact, even if we have built a package with conventional auxiliary functions (such as Gaussian type functions), it is possible that it deeply affects the behavior of certain solutions, leading to the problems encountered in the context of this model and others. In this sense, although Schutz formalism is not necessarily responsible for anomalous aspects of the operator that describes the quantum system, its use may cooperate to make the subject solution behavior changes during the procedure of obtaining the package.

The possibility that certain functions used in the construction of the wave packet can lead to the appearance of anomalies and physically inadequate results should be investigated more thoroughly in future work. An alternative is to compare solutions based on different wave packets. Other perspectives are the construction of models with other types of fluid or even the introduction of a second fluid (though this is a somewhat artificial process), which would further increase the number of degrees of freedom of the minisuperspace. This may allow a deeper understanding of the connection between matter and geometry in the early universe.

\vspace{1cm}

\textbf{Acknowledgments:} This work has received partial financial supporting from CNPq (Brazil), CAPES (Brazil) and FAPES (Brazil). Special thanks are due to P.V. Moniz and N.A. Lemos for discussions on the manuscript.

\end{document}